\documentclass[twocolumn,amsmath,amssymb,prl]{revtex4}

\usepackage{epsfig}

\def\al{\alpha}
\def\be{\beta}
\def\ga{\gamma}
\def\de{\delta}
\def\ep{\epsilon}

\def\ze{\zeta}
\def\et{\eta}

\def\ka{\kappa}
\def\la{\lambda}

\def\rh{\rho}

\def\ch{\chi}

\def\om{\omega}

\def\La{\Lambda}

\def\cG{{\cal G}}
\def\cl{{\cal L}}

\def\fr#1#2{{{#1} \over {#2}}}
\def\half{{\textstyle{1\over 2}}}

\def\frac#1#2{{\textstyle{{#1}\over {#2}}}}

\def\lsim{\mathrel{\rlap{\lower4pt\hbox{\hskip1pt$\sim$}}
    \raise1pt\hbox{$<$}}}
\def\gsim{\mathrel{\rlap{\lower4pt\hbox{\hskip1pt$\sim$}}
    \raise1pt\hbox{$>$}}}

\def\prt{\partial}

\def\etal{{\it et al.}}
\def\pt#1{\phantom{#1}}

\newcommand{\beq}{\begin{equation}}
\newcommand{\eeq}{\end{equation}}
\newcommand{\bea}{\begin{eqnarray}}
\newcommand{\eea}{\end{eqnarray}}
\newcommand{\bit}{\begin{itemize}}
\newcommand{\eit}{\end{itemize}}
\newcommand{\rf}[1]{(\ref{#1})}

\def\kr{{(k^{(4)})}{}}
\def\kdr{{(k^{(5)})}{}}
\def\kddr{{(k^{(6)}_1)}{}}
\def\krr{{(k^{(6)}_2)}{}}

\def\kb{\overline{k}{}}
\def\kt{\widetilde{k}{}}

\def\kbddr{{(\kb^{(6)}_1)}{}}
\def\kbrr{{(\kb^{(6)}_2)}{}}

\def\tb{\overline{t}}

\def\sb{\overline{s}}

\def\sbddr{{(\sb^{(6)}_1)}{}}
\def\sbh{\widehat{\overline{s}}{}}

\def\ub{\overline{u}}

\def\ubddr{{(\ub^{(6)}_1)}{}}
\def\ubh{\widehat{\overline{u}}}

\def\kl{{\ka\la}}
\def\mn{{\mu\nu}}
\def\ab{{\al\be}}
\def\abl{{\al\be\ldots}}
\def\abgd{{\al\be\ga\de}}
\def\abkl{{\al\be\ka\la}}
\def\gdmn{{\ga\de\mu\nu}}
\def\klmn{{\ka\la\mu\nu}}
\def\ol#1{\overline{#1}}

\def\cld#1{\cl^{(#1)}_{\rm LV}}
\def\cG{{G}}

\def\mbf#1{\boldsymbol #1}

\begin{document}

\title{Short-range gravity and Lorentz violation}

\author{Quentin G.\ Bailey$^1$, V.\ Alan Kosteleck\'y$^2$, and Rui Xu$^2$}

\affiliation{
$^1$Physics Department, Embry-Riddle Aeronautical University,
Prescott, AZ 86301, U.S.A.\\
$^2$Physics Department, Indiana University,
Bloomington, IN 47405, U.S.A.}

\date{IUHET 587, October 2014;
published as Phys.\ Rev.\ D {\bf 91}, 022006 (2015)}

\begin{abstract}

Comparatively few searches have been performed
for violations of local Lorentz invariance 
in the pure-gravity sector.
We show that tests of short-range gravity
are sensitive to a broad class of unconstrained and novel signals 
that depend on the experimental geometry and on sidereal time.

\end{abstract}

\maketitle

Gravity is a universal but comparatively weak force.
This makes it challenging to study and today,
some 350 years after Newton's {\it Principia},
our experimental understanding of gravity 
remains in some respects remarkably limited.
On the scale of the solar system,
we are confident that Newton's law describes the dominant physics
and that Einstein's General Relativity 
provides accurate relativistic corrections.
However,
on larger scales we lack a complete and compelling understanding,
as evidenced by dark energy.
On smaller scales below about 10 microns,
it is presently unknown whether gravity obeys Newton's law,
and forces vastly stronger than the usual inverse-square behavior 
remain within the realm of possibility. 

Perhaps the most crucial founding principle of General Relativity
is the Einstein equivalence principle.
Two of its ingredients are the weak equivalence principle,
which essentially states that gravity is flavor independent,
and local Lorentz invariance,
which states that rotations and boosts are local symmetries of nature.
Developing a deep understanding of gravity at all scales 
therefore depends on strong experimental support for these principles. 
The weak equivalence principle has been widely tested,
but tests of local Lorentz invariance have been 
largely limited to the pure-matter sector or to matter-gravity couplings
\cite{tables,cw}.
Here,
we undertake to address this gap 
by focusing on violations of local Lorentz symmetry
in the pure-gravity sector.

Effective field theory is a powerful and unique tool
for investigating physics at attainable scales
when definitive knowledge of the underlying physics is lacking.
It is therefore well suited for exploration
of local Lorentz invariance in gravity.
Indeed,
the pure-gravity sector
of the effective field theory describing
general local Lorentz violations for spacetime-based gravitation 
can be formulated as a Lagrange density 
containing the usual Einstein-Hilbert term and cosmological constant,
together with an infinite series of operators 
of increasing mass dimension $d$
representing corrections to known physics
at attainable scales
\cite{akgrav}.
To date,
however,
experimental searches for local Lorentz violation 
\cite{2007Battat,2007MullerInterf,2009Chung,%
2010Bennett,2012Iorio,2013Bailey,2014Shao}
and phenomenological studies
\cite{bk,ms}
within this framework
have been restricted to the so-called minimal sector,
consisting of terms with operators of the lowest mass dimension $d=4$.

In the present work,
we initiate a systematic study of local Lorentz violation with $d>4$,
introducing explicit expressions for $d=5$ and 6
and investigating prospective experimental constraints.
Operators of higher mass dimension $d$ involve more derivatives,
which translate to corrections to the Newton force law 
varying as $1/r^{d-2}$.
Short-range tests of gravity therefore offer
the sharpest sensitivities to effects from operators with $d>4$
and are our focus in what follows.
Moreover,
as discussed below,
the predicted signals contain novel features
that to date are unexplored in experiments.

We focus here on spontaneous violation of Lorentz symmetry
\cite{ksp}
in spacetime theories of gravity,
since the alternative of explicit Lorentz violation 
is generically incompatible with conventional Riemann geometry
or is technically unnatural in such theories 
\cite{akgrav}.
Spontaneous Lorentz violation occurs 
when an underlying action with local Lorentz invariance
involves gravitational couplings to tensor fields $k_\abl$
that acquire nonzero background values $\kb_\abl$
\cite{fn1}.
The field fluctuations $\kt_\abl\equiv k_\abl - \kb_\abl$
include massless Nambu-Goldstone and massive modes that affect the physics. 
The presence of nonzero backgrounds means 
the resulting gravitational phenomenology violates local Lorentz invariance,
and so the backgrounds $\kb_\abl$ 
are called coefficients for Lorentz violation
\cite{sme}.

In typical post-newtonian applications,
the coefficients $\kb_\abl$ are assumed small on the relevant physical scale
and constant in asymptotically flat coordinates,
and the analysis is performed
at linear order in the metric fluctuation $h_\ab$ 
and the coefficients $\kb_\abl$. 
Elimination of the fluctuations $\kt_\abl$ can be achieved 
by imposing the underlying diffeomorphism invariance on the dynamics,
thereby yielding a modified Einstein equation
expressed in terms of $\kb_\abl$
and quantities such as the linearized curvature tensor
\cite{bk}.
The phenomenology of the modified equation
can then be explored and experimental studies performed
to search for local Lorentz violation.

More explicitly,
we can write the Lagrange density of the underlying action
as the sum of four terms,
\beq
\cl = \cl_{\rm EH} + \cl_{\rm LV} + \cl_k + \cl_{\rm M},
\eeq
where $\cl_{\rm EH}= \sqrt{-g}(R - 2 \La)/16\pi G_N$
is the usual Einstein-Hilbert term with cosmological constant $\La$,
$\cl_{\rm LV}$ describes the gravitational coupling to the coefficient fields 
and hence is the source of phenomenological gravitational Lorentz violation,
$\cl_k$ contains the dynamics of the coefficient fields 
triggering the spontaneous Lorentz violation,
and $\cl_{\rm M}$ describes the matter.
The term $\cl_{\rm LV}$ 
can be written as a series 
involving covariant gravitational operators of increasing mass dimension $d$,
\beq
\cl_{\rm LV} = \fr {\sqrt{-g}}{16\pi G_N}
(\cld4 +\cld5 +\cld6 + \ldots ).
\eeq
Each term is formed by contracting the coefficient fields $k_\abl$
with gravitational quantities 
including covariant derivatives $D_\al$ and curvature tensors $R_{\abgd}$.
Here,
we consider explicitly terms with $4\leq d\leq 6$,
though much of our discussion can be directly generalized to larger $d$.

The minimal term $\cld4$ with $d=4$ is 
\cite{akgrav} 
\beq
\cld4 = \kr_{\abgd} R^{\abgd}.
\label{minimallag}
\eeq
The dimensionless coefficient field $\kr_{\abgd}$
inherits the symmetries of the Riemann tensor
and can be decomposed into its traceless part 
$t_{\abgd}$,
its trace $s_{\ab}$,
and its double trace $u$.
Within the post-newtonian treatment outlined above,
the coefficient $\ub$ acts as an unobservable rescaling of $G_N$
\cite{fn2}.
In pure gravity, 
the coefficient $\sb_{\ab}$ can be removed via coordinate definitions
\cite{akgrav},
but more generally it generates many phenomenological effects,
and its 9 independent components have been constrained 
to varying degrees down to about $10^{-10}$
by numerous analyses using data from
lunar laser ranging
\cite{2007Battat},
atom interferometry
\cite{2007MullerInterf,2009Chung},
short-range tests
\cite{2010Bennett},
satellite ranging
\cite{2012Iorio},
precession of orbiting gyroscopes
\cite{2013Bailey},
pulsar timing
\cite{2014Shao},
and perihelion and solar-spin precession
\cite{bk,2012Iorio}.
The coefficient $\tb_{\abgd}$ is absent
at leading orders in the post-newtonian expansion,
and to date its 10 independent components have no known physical implications
for reasons that remain mysterious (the `$t$ puzzle').

For $d=5$,
the general expression using curvature and covariant derivatives is 
\beq
\cld5 = \kdr_{\abgd\ka} D^\ka R^{\abgd}.
\eeq 
In the linearized limit,
or more generally under the operational definition
of the CPT transformation
\cite{akgrav},
the expression $D^\ka R^{\abgd}$ is CPT odd.
Any effects from $\cld5$ in the nonrelativistic limit
would therefore represent 
pseudovector contributions to the associated Newton gravitational force 
rather than conventional vector ones,
and hence they would lead to self accelerations of localized bodies.
Analogous issues are known for some CPT-odd terms in other sectors
\cite{km09}.
Any stable models with terms of the form $\cld5$
therefore cannot lead to effects on nonrelativistic gravity,
and so their phenomenology lies outside our present scope.

Instead, we focus on Lorentz violation at $d=6$, 
for which we write $\cld6$ in the form
\bea
\cld6 &=& \half \kddr_{\abgd\kl} \{D^\ka, D^\la\}  R^{\abgd} 
\nonumber\\
&&
+ \krr_{\abgd\klmn} R^{\klmn} R^{\abgd}.
\label{lvlag}
\eea
The coefficient fields $\kddr_{\abgd\kl}$ and $\krr_{\abgd\klmn}$
have dimensions of squared length,
or squared inverse mass in natural units.
In the first term,
the anticommutator of covariant derivatives suffices for generality
because including the commutator 
would merely duplicate part of the second term.
The first four indices on $\kddr_{\abgd\kl}$ 
inherit the symmetries of the Riemann tensor,
as do the first and last four indices on $\krr_{\abgd\klmn}$, 
while the Bianchi identity implies 
the additional cyclic-sum condition
$\sum_{(\ga\de\ka)} \kddr_{\abgd\kl} = 0$.
The number of independent components 
in $\kddr_{\abgd\kl}$ and $\krr_{\abgd\klmn}$
is therefore 126 and 210, respectively. 
The coefficients 
$\kddr_{\abgd\kl}$ 
could arise from Lorentz-violating derivative couplings 
of fields to gravity in the underlying theory.
Models of this type are straightforward to construct,
although we are unaware of examples in the literature.
The coefficients 
$\krr_{\abgd\klmn}$
represent general quadratic Lorentz-violating curvature couplings,
specific forms of which occur in many models 
as a result of integrating over fields in the underlying action 
that have Lorentz-violating couplings to gravity.
Examples include 
Chern-Simons gravity
\cite{jp,ay},
the cardinal model
\cite{cardinal},
and various types of bumblebee models
\cite{bumblebee1,akgrav,bumblebee2}.
 
To extract the linearized modified Einstein equation
resulting from the terms \rf{lvlag},
we assume an asymptotically flat background metric $\et_\ab$ as usual, 
and write the background coefficients as
$\kbddr_{\abgd\kl}$ and $\kbrr_{\abgd\klmn}$.
We remark that the procedure 
for linearization and elimination 
of coefficient fluctuations outlined above 
\cite{bk}
involves no fluctuations for $\krr_{\abgd\klmn}$
because these contribute only at nonlinear order.
After some calculation,
we find the linearized modified Einstein equation
can be written in the form
\bea
G_\mn &=& 8\pi G_N (T_M)_\mn 
+2\sbh^{\al\be} \cG_{\al(\mu\nu)\be}
-\frac 12 \ubh G_\mn
\nonumber\\
&&
+a \kbddr_{\al(\mu\nu)\be\ga\de} \prt^\al \prt^\be R^{\ga\de}
\nonumber\\
&&
+ 4 \kbrr_{\al\mu\nu\be\ga\de\ep\ze} \prt^\al \prt^\be R^{\ga\de\ep\ze},
\label{lineq3}
\eea
where
$\cG_\abgd \equiv \ep_\abkl \ep_\gdmn R^\klmn/4$
is the double dual of the Riemann tensor
and $G_{\al\be} \equiv \cG^{\ga}_{\pt{\ga}\al\ga\be}$ is the Einstein tensor.
In Eq.\ \rf{lineq3},
all gravitational tensors are understood to be linearized in $h_\mn$.
Also,
we have introduced the scalar operator
$\ubh = \ub + \ubddr_{\al\be} \prt^\al \prt^\be$
and the tensor operator 
$\sbh_{\al\be} = \sb_{\al\be} + \sbddr_{\abgd} \prt^\ga \prt^\de$,
where
$\ubddr_{\ga\de} \equiv \kbddr^{\al\be}_{\pt{\al\be}\abgd}$
is a double trace and 
$\sbddr^{\al}_{\pt{\al}\be\ga\de} \equiv 
\kbddr^{\al\ep}_{\pt{\al\ep}\be\ep\ga\de} 
- \de^\al_{\pt{\be}\be} \ubddr_{\ga\de}/4 $
involves a single trace.
Note that the entire contribution 
from the $d=4$ Lorentz-violating term \rf{minimallag}
is contained in $\ubh$ and $\sbh_{\al\be}$,
along with comparable pieces of the $d=6$ derivative term.
This structure may offer some insight into the $t$ puzzle mentioned above.
The parameter $a$ in Eq.\ \rf{lineq3}
is a model-dependent real number
that depends on the dynamics specified by the Lagrange density $\cl_k$.

The modified Einstein equation \rf{lineq3}
is likely to imply numerous phenomenological consequences
both for relativistic effects such as gravitational waves
and for nonrelativistic effects in post-newtonian gravity.
Here,
we consider the nonrelativistic limit with zero cosmological constant
and for an extended source with mass density $\rh(\mbf r)$.
The modified Einstein equation for the $d=6$ terms then 
reduces to a modified Poisson equation of the form
\beq
-\vec \nabla^2 U = 
4\pi G_N \rh 
+ (\kb_{\rm eff})_{jklm} \prt_j \prt_k \prt_l \prt_m U,
\label{newt}
\eeq
where $U(\mbf r)$ is the modified Newton gravitational potential.
In this equation,
$(\kb_{\rm eff})_{jklm}$ are effective coefficients for Lorentz violation 
with totally symmetric indices,
revealing that the number of independent observables 
for Lorentz violation in the nonrelativistic limit is 15.
These effective coefficients are linear combinations 
of the $d=6$ coefficients 
$\kddr_{\abgd\kl}$ and $\krr_{\abgd\klmn}$,
the explicit form of which is somewhat lengthy
and irrelevant for present purposes and so is omitted here,
but we remark in passing that 
many of the independent components
$\kddr_{\abgd\kl}$ and $\krr_{\abgd\klmn}$ appear.

To solve the modified Poisson equation \rf{newt}
we can adopt a perturbative approach,
with the Lorentz-violating term assumed to generate
a small correction to the usual Newton potential.
This is consistent with the notion 
that the $d=6$ Lorentz-violating term \rf{lvlag}
represents a perturbative correction to the Einstein-Hilbert action 
on the length scales of experimental interest.
The nonperturbative scenario with $\cld6$ dominating the physics
could in principle also be of interest
but involves theoretical complexities that lie outside our present scope. 
Within the perturbative assumption, 
the solution to the modified Poisson equation 
\rf{newt}
can be written as
\bea
U (\mbf r) &=&  
G_N \int d^3 r^\prime 
\fr{\rh (\mbf r^\prime )}{ |\mbf r - \mbf r^\prime|}
\left( 1 + \fr { \kb (\widehat{\mbf R}) } {
|\mbf r - \mbf r^\prime|^2} \right) 
\nonumber\\
&&
+{\frac 45 \pi G_N} \rh(\mbf r) (\kb_{\rm eff})_{jkjk},
\label{ULV2}
\eea
where $\widehat {\mbf R}
= (\mbf r - \mbf r^\prime)/
|\mbf r - \mbf r^\prime|$.
The quantity $\kb = \kb (\hat{\mbf r})$
is an anisotropic combination of coefficients 
and a function of $\hat{\mbf r}$,
given by
\bea
\kb (\widehat{\mbf r}) &=& \frac 32  (\kb_{\rm eff})_{jkjk}
- 9  (\kb_{\rm eff})_{jkll} \hat{r}^j \hat{r}^k 
\nonumber\\
&&
+ \frac {15}{2} (\kb_{\rm eff})_{jklm} \hat{r}^j \hat{r}^k \hat{r}^l \hat{r}^m.
\label{tildekb}
\eea
The potential \rf{ULV2} 
contains the conventional Newton potential
and a correction term that varies with the inverse cube of the distance.
The final piece is a contact term
that becomes a delta function in the point-particle limit,
in parallel with the usual dipole contact term in electrodynamics. 

The inverse-cube behavior of the potential
leads to an inverse-quartic gravitational field 
$\mbf g = \mbf\nabla U$.
The rapid growth of the force at small distances
suggests that the best sensitivities 
to Lorentz violation
could be achieved in experiments on short-range gravity
\cite{review},
which measure the deviation from the Newton gravitational force
between two masses.
Next,
we consider the signals in experiments of this type.

In an Earth-based laboratory,
measurements of the force between two masses 
are instantaneously sensitive to the coefficients 
$(\kb_{\rm eff})_{jklm}$ in the local frame.
However,
the laboratory frame is noninertial,
so the Earth's rotation about its axis
and revolution about the Sun
induce variations of these coefficients with sidereal time $T$.
The canonical frame adopted for reporting results
from experimental searches for Lorentz violation 
is the Sun-centered frame
\cite{tables,sunframe},
with $Z$ axis along the direction of the Earth's rotation
and $X$ axis pointing towards the vernal equinox 2000.
Neglecting the Earth's boost,
which is of order $10^{-4}$,
the transformation from the Sun-centered frame $(X,Y,Z)$
to the laboratory frame $(x,y,z)$ can be accomplished 
using a time-dependent rotation $R^{jJ}$,
where $j = x,y,z$ and $J=X,Y,Z$.
For example,
taking the laboratory $z$ axis pointing to the local zenith
and the $x$ axis pointing to local south,
the rotation matrix is 
\beq
R^{jJ}=\left(
\begin{array}{ccc}
\cos\ch\cos\om_\oplus T
&
\cos\ch\sin\om_\oplus T
&
-\sin\ch
\\
-\sin\om_\oplus T
&
\cos\om_\oplus T
&
0
\\
\sin\ch\cos\om_\oplus T
&
\sin\ch\sin\om_\oplus T
&
\cos\ch
\end{array}
\right),
\label{rotmat}
\eeq
where the angle $\ch$ is the colatitude of the laboratory
and $\om_\oplus\simeq 2\pi/(23{\rm ~h} ~56{\rm ~min})$ 
is the Earth's sidereal frequency.
The $T$-dependent coefficients $(\kb_{\rm eff})_{jklm}$ 
in the laboratory frame are then given by
\beq
(\kb_{\rm eff})_{jklm} = R^{jJ} R^{kK} R^{lL} R^{mM} (\kb_{\rm eff})_{JKLM}
\label{rot}
\eeq
in terms of constant coefficients $(\kb_{\rm eff})_{JKLM}$ 
in the Sun-centered frame.

The sidereal variation of the laboratory-frame coefficients
implies that the modified potential $U$ and force 
between two masses measured in the laboratory frame 
vary with time $T$. 
For example,
the modified potential due to a point mass $M$ takes the form
\beq
U (\mbf r,T) =  \fr {G_N M}{r} 
\left( 1 + \fr {\kb (\hat {\mbf r},T)}{r^2} \right)
\label{ULV3}
\eeq
away from the origin,
where Eq.\ \rf{rot} is used to express
the combination ${\kb (\hat {\mbf r},T)}$
in Eq.\ \rf{tildekb}
in terms of coefficients $(\kb_{\rm eff})_{JKLM}$ 
in the Sun-centered frame.
The modified force therefore depends
both on direction and on sidereal time,
which leads to striking signals in short-range experiments.
For example,
the time dependence in Eq.\ \rf{rot} 
implies that the effective gravitational force between two bodies 
can be expected to vary with frequencies 
up to and including the fourth harmonic of $\om_\oplus$.
Also,
the direction dependence 
of the laboratory-frame coefficients $(\kb_{\rm eff})_{jklm}$
implies an asymmetric dependence of the signal on the shape of the bodies.
A few simple results valid in conventional Newton gravity,
such as the constancy of the force at any point above an infinite plane
of uniform mass density,
still hold for the potential \rf{ULV3}.
However,
for the finite bodies used in experiments
it is typically necessary to determine the potential and force
via numerical integration.
Indeed,
simple simulations for experimental configurations such as 
two finite planes 
\cite{colorado03}
or a plane and a sphere 
\cite{iupui03}
reveal that shape and edge effects
play an important role in determining the sensitivity of the experiment
to the coefficients for Lorentz violation.
 
The modified potential \rf{ULV3} involves an inverse-cube correction
to the usual Newton result.
Its time and orientation dependence means that existing experimental limits 
on spherically symmetric inverse-cube potentials 
cannot be immediately converted into constraints 
on the coefficients $(\kb_{\rm eff})_{JKLM}$,
as typical experiments collect data over an extended period 
and disregard the possibility of orientation-dependent effects. 
Establishing definitive constraints on the coefficients
$(\kb_{\rm eff})_{JKLM}$ for Lorentz violation
will therefore require new experimental analyses.
Next,
we illustrate some of the issues involved
by considering briefly one particular example:
the E\"otWash limit on inverse-cube potentials
obtained using a torsion pendulum
\cite{eotwash07,inversesquare}.

The apparatus in this experiment 
consists of a test-mass bob in the shape of a disk with 42 cylindrical holes
arranged in two concentric circles,
suspended by a fiber through its center and normal to its plane.
A similar disk serving as the source mass is placed below and rotated,
thereby producing a periodic torque on the upper disk
of strength and harmonic signature 
determined by deviations from the inverse square law.
The experiment yielded a limit 
\cite{inversesquare}
on a spatially homogeneous and time-independent inverse-cube potential
that in the present context can be interpreted as
a constraint on an averaged coefficient given by
$\langle{\kb (\hat {\mbf r},T)}\rangle <1.3\times 10^{-10} {\rm ~m}^2$
at the $68\%$ confidence level.
The averaging involves both spatial and time smearing,
which cannot be performed exactly 
without careful modeling of the apparatus
and incorporating the time stamps for the data. 
Nonetheless,
a crude estimate for the type of constraint 
that would emerge from a detailed reanalysis
can be obtained by modeling the apparatus
using a numerical simulation involving 21 point masses on a ring
above another 21 point masses on a second ring rotating at fixed frequency.
Using the transformation \rf{rotmat} for colatitude $\ch\simeq 42^\circ$ 
and averaging the results over a sidereal day
reveals that in this simple simulation
only six independent coefficients 
control the averaged Lorentz-violating torque,
and they appear in the combination
\bea
\ol k_{\rm simulation} &\equiv& 
(\ol k_{\rm eff})_{XXZZ} + (\ol k_{\rm eff})_{YYZZ} 
\nonumber\\
&&
+ 0.4(\ol k_{\rm eff})_{XXXX} + 0.4(\ol k_{\rm eff})_{YYYY}
\nonumber\\
&&
+ 0.8 (\ol k_{\rm eff})_{XXYY} + 0.3(\ol k_{\rm eff})_{ZZZZ} .
\quad
\label{simul}
\eea
As expected for an averaging analysis,
the torque is found to mimic closely that obtained
using a spherically symmetric inverse-cube potential.
Using Eqs.\ \rf{tildekb} and \rf{rot}
together with the above experimental constraint on
$\langle{\kb (\hat {\mbf r},T)}\rangle$,
we can deduce the crude constraint 
$|\ol k_{\rm simulation}| \lsim  10^{-11} {\rm ~m}^2$.
Although only an approximation to an exact analysis,
this procedure does give a feel for the sensitivity to Lorentz violation
currently attainable in tests of short-range gravity.

Given the novel features 
of short-range tests of local Lorentz violation in gravity
and the wide variety of experiments in the literature,
it is useful to identify a measure
serving as a rapid gauge of the reach of a given experiment.
As seen above,
a definitive answer to this question requires
careful simulation of the experiment,
but a rough estimate can be obtained by taking advantage
of the common practice for experiments testing short-range gravity
to report results in terms of two parameters $\al$, $\la$
appearing in a potential modified by a Yukawa-like term, 
$U_{\rm Yukawa} = {G_N M} ( 1+ \al e^{-r/\la})/r$.
Comparing this Yukawa form with the potential \rf{ULV3} 
indicates that experiments attaining sensitivities 
to $\al$ and $\la$ at distances $r\approx \la$
can be expected to have sensitivities to Lorentz violation of order 
$|{\kb (\hat {\mbf r},T)}| \approx \al \la^2/e$
and hence using Eq.\ \rf{tildekb}
a coefficient reach of order 
\beq
|(\kb_{\rm eff})_{JKLM}| \approx \al \la^2/10.
\label{limit}
\eeq
Note,
however,
that sensitivity to the perturbative Lorentz violation considered here 
implies that the experiment must be able to detect usual Newton gravity,
which is the case for only a subset of experiments
reported in the literature.
Note also that different experiments are typically sensitive
to distinct linear combinations of $(\kb_{\rm eff})_{JKLM}$.

Within this perspective,
the most interesting short-range experiments
are those at small $\la$ that are sensitive to the usual Newton force.
For example,
the E\"otWash experiment described above achieves sensitivity of order 
$\al \simeq 10^{-2}$ at $\la \simeq 10^{-4}$ m,
which suggests a reach for Lorentz violation of order 
$|(\kb_{\rm eff})_{JKLM}| \simeq 10^{-11}$ m$^{2}$,
in agreement with the estimate from the simulation \rf{simul}.
As another example,
the Wuhan experiment 
\cite{hust12}
attains 
$\al \simeq 10^{-3}$ at $\la \simeq 10^{-3}$ m,
for which Eq.\ \rf{limit} gives the estimate 
$|(\kb_{\rm eff})_{JKLM}| \simeq 10^{-10}$ m$^{2}$.
Similarly,
the early Irvine experiment
\cite{irvine85}
achieved $\al \simeq 3\times 10^{-3}$ at $\la \simeq 10^{-2}$ m,
yielding an approximate reach of order 
$|(\kb_{\rm eff})_{JKLM}| \simeq 3 \times 10^{-8}$ m$^{2}$.
In contrast,
the Indiana experiment
\cite{colorado03}
sits on the cusp of the perturbative limit,
achieving
$\al \simeq 1$ at $\la \simeq 10^{-4}$ m
and hence having an estimated sensitivity of order 
$|(\kb_{\rm eff})_{JKLM}| \simeq 10^{-9}$ m$^{2}$.

In some gravity theories with violations of Lorentz invariance,
the predicted effects can be comparatively large 
while escaping detection to date
\cite{kt}.
The above estimated sensitivities
suggest terms in the pure-gravity sector with $d>4$
are interesting candidates for such countershaded effects
because the Planck length $\simeq 10^{-35}$ m lies far below
the range accessible to existing laboratory experiments on gravity.
In any case,
short-range tests of gravity offer an excellent opportunity
to search for local Lorentz violation involving operators with $d>4$,
thereby establishing the Einstein equivalence principle
for the pure-gravity sector on a complete and firm experimental footing.

We thank Ricardo Decca and Josh Long for valuable discussions.
This work was supported in part
by the Department of Energy
under grant number {DE}-SC0010120,
by the National Science Foundation
under grant number PHY-1402890,
and by the Indiana University Center for Spacetime Symmetries.

\end{document}